\newcommand{\bra}[1]{\langle {#1} |}
\newcommand{\ket}[1]{| {#1} \rangle }
\newcommand{\braket}[2]{\langle {#1} | {#2} \rangle }

\newcommand{\Tr}{{\mathrm{Tr}}}
\newcommand{\half}{{\frac{1}{2}}}
\documentclass{article}
\usepackage{epsfig}

\usepackage{amsmath}
\usepackage{graphics}

\begin{document}

\setlength{\textheight}{8.0truein}    %FOR 2ND PAGE ONWARDS

%\runninghead{Fishing for Eavesdroppers}
            %{P.~Lopata and T.~Bahder}

%\normalsize\textlineskip
\thispagestyle{empty}
\setcounter{page}{1}

%\copyrightheading{Vol.}{No.}{Year}{Page Nos.}
%\copyrightheading{0}{0}{2003}{000--000}

\vspace*{0.88truein}

%\alphfootnote

%\fpage{1}

\centerline{\bf
%%%%%%%%%%%%%%%%%%%%%
%Put in titiles here
%%%%%%%%%%%%%%%%%%%%%
FISHING FOR EAVESDROPPERS}

\vspace*{0.37truein}
\centerline{\footnotesize
%%%%%%%%%%%%%%%%%%%%%%%%%%%%%%%%%%%%
%put authors' name and address here
%%%%%%%%%%%%%%%%%%%%%%%%%%%%%%%%%%%%
PAUL A. LOPATA%\footnote{current address:}
}
\vspace*{0.015truein}
\centerline{\footnotesize\it U.S.~Army Research Laboratory} 
\baselineskip=10pt
\centerline{\footnotesize\it 2800 Powder Mill Road, Adelphi, MD 20783 USA}
\vspace*{10pt}
\centerline{\footnotesize 
THOMAS B. BAHDER}
\vspace*{0.015truein}
\centerline{\footnotesize\it Charles M.~Bowden Research Facility}
\baselineskip=10pt
\centerline{\footnotesize\it Aviation and Missile Research, Development and Engineering Center}
\baselineskip=10pt
\centerline{\footnotesize\it US Army RDECOM, Redstone Arsenal, Alabama, 35898 USA}
\vspace*{0.225truein}
%\publisher{(received date)}{(revised date)}

\vspace*{0.21truein}

%% \abstracts{first paragraph}{second paragraph}{third paragraph}
%% If there is only one paragraph, just keep the second and third empty 
%% like the following one 
\abstract{
%%%%%%%%%%%%%%%%%%%%
% put abstract here
%%%%%%%%%%%%%%%%%%%%
A method is given to detect the presence of eavesdroppers when a noisy message is sent to a privileged receiver.   A proof of the effectiveness if this method is demonstrated, and a comparison is made to other quantum cryptographic tasks.  
}{}{}

\vspace*{10pt}

%\keywords{quantum cryptography, quantum information}
\vspace*{3pt}
%\communicate{to be filled by the Editorial}

\vspace*{1pt}%\textlineskip    %) USE THIS MEASUREMENT WHEN THERE IS
   %) A SECTION HEADING
%\vspace*{-0.5pt}
%\noindent
%%%%%%%%%%%%%%%%%%%%%%%%%%%%%%%%
%put the text of the paper here
%%%%%%%%%%%%%%%%%%%%%%%%%%%%%%%%

%\maketitle
\section{Introductory Remarks}
The aim of this work is to introduce a method for accomplishing the following cryptographic task: the transmission of a message from its sender to a privileged receiver so that the receiver can detect the presence of any eavesdroppers during the transmission.  Specifically, this method provides the receiver with a confirmation that no eavesdroppers were present provided that this is indeed the case and also provided that there was a small enough amount of noise present in the channel they were using to communicate.  In other words, an eavesdropper will be indistinguishable from noise, but the absence of any eavesdropping and noise below a certain threshold will allow the message receiver to have confidence that the message that was just transferred had not been exposed to anyone other than himself.  

It is important to emphasize that this method does not prevent any third parties from gaining access to the message.  Indeed, it should be clearly understood that messages sent using this method can be read by an eavesdropper (we describe the most effective technique for doing so below).  However, any such activity will be detectable by the message receiver, as we shall demonstrate.

This manuscript is organized as follows: In the next section, the method for achieving the cryptographic goal just described is explained.  Following this, we give a mathematical proof  that this method allows the message receiver to determine if a message has not been exposed to any eavesdroppers.  In the final section we compare this cryptographic task to other cryptographic activities.

\section{Description of the Protocol}
The protocol involves a message sender, named Alice, trying to get a message bit $b$ (whose value can be zero or one) to a receiver, called Bob.   Longer messages will be built up by sending a sequence of message bits.  The entire process is initiated by Alice announcing that she has a message bit she wants to communicate to Bob.  Then a routine is repeated many times.  Each repetition is referred to as a single \textit{shot} and proceeds as follows.    

Bob starts the routine by preparing some physical system, which we refer to as a particle, in one of four quantum states represented by density matrices $\rho_0 \equiv \ket{0}\bra{0}$, $\rho_1 \equiv \ket{1}\bra{1}$, $\rho_+ \equiv \ket{+}\bra{+}$, and $\rho_- \equiv \ket{-}\bra{-}$.  Here we have used the standard notation for a qubit Hilbert space that is spanned by orthonormal basis vectors $\ket{0}$ and $\ket{1}$, and $\ket{\pm} \equiv \bigl(\ket{0}\pm \ket{1}\bigr)/\sqrt{2}$.  Bob employs a randomized method of preparation so that he is equally likely to prepare the particle in any of these four states.  Bob records which state he has prepared and sends the particle to Alice.  

Upon receipt of the particle, Alice makes one of two measurements: with probability one-half she makes a measurement corresponding to $\sigma_1 \equiv \ket{+}\bra{+} - \ket{-}\bra{-}$, and with probability one-half she makes a measurement corresponding to $\sigma_3\equiv \ket{0}\bra{0} - \ket{1}\bra{1}$.  Alice records her measurement result as $m$ (which will be $+1$ or $-1$) and announces whether her measurement corresponded to $\sigma_1$ or $\sigma_3$.    

Then Alice makes a second announcement.  With probability $p_a$ she announces the classical bit $a$ which is determined from the message bit $b$ and the measurement result $m$ using 
\begin{equation}\label{mod2} 
a = \left( b + \frac{1-m}{2} \right)\! \! \mod 2 \ ,
\end{equation}  
which we refer to as a \textit{bit-announcement}. Or else, (with probability $[1-p_a]$) she announces the value of $m$, which we refer to as a \textit{result-announcement}.  This completes the routine for a single shot.  

\begin{table} 
\caption{\label{AliceAnnouncements} Each shot, Alice makes one of the following eight announcements. }
\begin{center}
%\begin{ruledtabular}
\begin{tabular}{{|c|c|}}
\hline
\hspace{\stretch{1}} Bit-Announcements \hspace{\stretch{1}} & \hspace{\stretch{1}} Result-Announcements \hspace{\stretch{1}} \\
\hline 
$\sigma_1 \phantom{mm} a=0$ & $\sigma_1 \phantom{mm} m=+1$ \\
$\sigma_1 \phantom{mm} a=1$ & $\sigma_1 \phantom{mm} m=-1$ \\
$\sigma_3 \phantom{mm} a=0$ & $\sigma_3 \phantom{mm} m=+1$ \\
$\sigma_3 \phantom{mm} a=1$ & $\sigma_3 \phantom{mm} m=-1$ \\
\hline
\end{tabular}
\end{center}
%\end{ruledtabular}
\end{table}

This routine must be repeated $N$ times to accomplish the goal of conveying the value of the message bit $b$ to Bob while giving him the opportunity to determine if he has been the only person to whom this message bit has been conveyed.  For the purpose of simplicity of the analysis, we assume that the value of $N$ has been agreed upon beforehand between Alice and Bob. (It is straightforward to extend the analysis shown here to a situation in which Alice and Bob keep repeating shots until some well-defined goal has been achieved.)  Furthermore, we assume that the value of $p_a$ has been agreed upon beforehand as well.  The criteria for the choice of these two parameters is discussed below.

Each time that Alice makes a bit-announcement, Bob has a chance to determine the value of $b$. (As we shall see shortly, each bit-announcement also gives an eavesdropper an opportunity to infer the value of $b$.)  When Bob has prepared the particle in an eigenstate of the measurement operator corresponding to the measurement that Alice performs, we say that they have a \textit{matching basis}.  In such a case, Bob can use his conjectured value of $m$ (based on his knowledge of the state he prepared and the measurement made by Alice) along with the equation  
\begin{equation} 
b = \left( a + \frac{1-m}{2} \right)\! \! \mod 2 \ 
\end{equation}
to determine the value of $b$.  On any given shot, the probability that Bob and Alice have a matching basis and that Alice makes a bit-announcement is  $p_a/2$.  If a bit-announcement is made on a shot when Bob knows that he and Alice do not have a matching basis, then from Bob's point of view, either value of $m$ is equally likely, and therefore either value of $b$ is equally likely.  

However, as in any physical system, the way in which the events occur does not always match the ideal.  These non-ideal events are often characterized as noise and will effect whether or not Bob can correctly infer the message bit $b$ from Alice's announcements.  We finish this section with an analysis of how well Bob can infer the value of the message bit in the presence of noise.  

Ideally, the probability $\Pr (m|\sigma_l, \rho_i)$ of the measurement result $m$ when Bob prepares the particle in state $\rho_i$ and Alice makes a measurement corresponding to $\sigma_l$ will be one-half when Bob and Alice do not have a matching basis and will be either zero or one when Bob and Alice do have a matching basis:
\begin{align*}
1 & = \Pr (m\! = \! +1|\sigma_1, \rho_+) = \Pr (m\! = \! -1|\sigma_1, \rho_-) \\
  & = \Pr (m\! = \! +1|\sigma_3, \rho_0) = \Pr (m\! = \! -1|\sigma_3, \rho_1) \ ,\\
0 & = \Pr (m\! = \! -1|\sigma_1, \rho_+) = \Pr (m\! = \! +1|\sigma_1, \rho_-) \\
  & = \Pr (m\! = \! -1|\sigma_3, \rho_0) = \Pr (m\! = \! +1|\sigma_3, \rho_1) \ . 
\end{align*}
In other words, a particular outcome is expected when Bob and Alice have a matching basis.  But (when noise is present) these probabilities will deviate from the ideal values.  As such, we characterize any events that occur on a shot when Bob and Alice have a matching basis, but that are not predicted by these idealized probabilities, as a \textit{mismatch}.  For example, if Bob prepared the particle in the state $\rho_+$ and Alice made the measurement corresponding to $\sigma_1$ and found the result $m=-1$, this would be considered a mismatch.  It is easy to see how such a mismatch might lead Bob to infer the incorrect value for the message bit $b$, since the result he expected Alice to find was not the result that she found.  In light of these possible mismatches, it is still possible to quantitatively characterize what Bob expects to learn from the use of this protocol.  

To do this, we use the Shannon mutual information $I(B\!:\!C)$ between the random variable $B$ describing the message bit $b$ and the random variable $C$ describing the compound events corresponding to the data that Bob acquires on shots where Alice makes a bit-announcement.  We do not consider the result-announcements in this calculation because they have no dependence on the message bit.  

The random variable $B$ describes the possible values of the message bit $b$, which are $b=0$ and $b=1$, along with their probabilities, which are assumed to be $\Pr(b=0) = \Pr(b=1)=1/2$.  

The random variable $C$ describes the data that Bob acquires during shots where Alice makes a bit-announcement.  For each of these shots, Bob records the compound event $c = (\rho_i, \sigma_l, a)$ which consists of the state $\rho_i$ in which he prepared the particle,  the operator $\sigma_l$ corresponding to Alice's announced measurement, and the bit-announcement  $a$.  At the end of $N$ shots, Alice will have made $k$ bit-announcements (with $0 \leq k \leq N$) and Bob will have recorded a string of compound events $\mathbf{c} = (c_1, c_2, \ldots, c_k)$ where the event $c_1$ occurs during the shot with the first bit-announcement,  $c_2$ on the second, and so forth.  Because of the probabilistic nature of the protocol, the length $k$ of this string is not known beforehand. The probability $\Pr(k)$ that Alice makes $k$ bit-announcements can be calculated using the binomial distribution:
\[
\Pr(k) = \frac{N!}{k! (N-k)! } p_a^k (1-p_a)^{N-k} \ .
\]
There are sixteen different compound events that can occur on each shot -- four possible initial states, two possible measurements, and two possible bit-announcements. When Alice makes $k$ bit-announcements there are $16^k$ possible compound-event strings.   This leads to $\sum_{k=0}^N 16^k = \frac{16^{N+1}-1}{15}$ different possible compound-event strings.  The random variable $C$ describes these strings of possible compound events along with their probabilities.  Because each shot is an independent event, the probability that a particular string $\mathbf{c} = (c_1, \ldots, c_k)$ occurs is simply the product of the probabilities for each single-shot compound event to occur: $\Pr(\mathbf{c}) = \prod_{j=1}^N \Pr(c_j)$.  The probability of one of the individual events $\Pr (\rho_i, \sigma_l, a) = \Pr (a|\rho_i, \sigma_l)\Pr(\rho_i, \sigma_l)= \Pr (a|\rho_i, \sigma_l)\Pr(\rho_i) \Pr(\sigma_l) = \frac{1}{8}\Pr (a|\rho_i, \sigma_l)$ is dependent upon the value of $b$ and the value of the measurement result $m$.  From Bob's point of view $\Pr (a|\rho_i, \sigma_l) = \frac{1}{2}\bigl[ \Pr(m=+1| \rho_i, \sigma_l, b=0) + \Pr(m=-1| \rho_i, \sigma_l, b=1) \bigr] = 1/2$, independent of any changes to the state of the particle as it travels to Alice. 

With this explicit description of the random variables $B$ and $C$, we are now in a position to determine the mutual information between them,
\[
I(B\!:\!C) = \sum_{b=0}^1 \Pr(b) \sum_{\textbf{c}} \Pr(\mathbf{c}|b) \log \Pr(\mathbf{c}|b) - \sum_{\textbf{c}} \Pr(\mathbf{c}) \log \Pr(\mathbf{c})  
\]
where the sum over $\mathbf{c}$ indicates that the summation is made over all possible compound-event strings, and where $\Pr (\mathbf{c}|b)$ is the probability of the compound-event string $\mathbf{c}$ given a particular value of the message bit $b$.  These can be calculated using $\Pr (a\!=\!b, \rho_i, \sigma_l|b) = \frac{1}{8}\Pr(m\!=\!+1|\sigma_l, \rho_i)$ and $\Pr (a\!\neq \!b, \rho_i, \sigma_l|b) = \frac{1}{8}\Pr(m\!=\!-1|\sigma_l, \rho_i)$ along with $\Pr(\mathbf{c}|b) = \prod_{j=1}^k\Pr(c_j|b)$.  Table \ref{conditional_probs} lists the various outcomes along with the probabilities that show up in the mutual information which depend upon the values of $D_1$, $D_3$, $D_{0+}$, and $D_{+0}$ which all fall within the range of zero to one.   With the ideal system discussed earlier, $D_1 = D_3 = 0$ and $D_{+0} = D_{0+}= 1/2$.  While the values of $D_1$ and $D_3$ will likely deviate from zero, it should be possible for Bob and Alice to calibrate their equipment properly to ensure that $D_{+0} = D_{0+} = 1/2$ in the absence of any outside interference with the system. If these two values deviate from $1/2$ then there is some systematic error with their alignment or there is something interacting with the particles in a non-symmetric way as they travel from Bob to Alice.  Therefore, we assume that it is possible for Bob and Alice to improve their apparatus until $D_{+0} = D_{0+} = 1/2$.  Furthermore, for similar symmetry reasons, we assume that Alice and Bob can ``rotate'' their systems until $D_1 = D_3$, and we call this value  $D = D_1 = D_3$.  In this way, these probabilities can be characterized by a single value, $D$.  Moreover, this value of $D$ is simply the probability of mismatch.  

\begin{table} 
\caption{\label{conditional_probs} The probabilities $\Pr (m|\sigma_l, \rho_i)$ that Alice's measurement corresponding to $\sigma_l$ results in the value $m$ given that Bob prepared the particle in the state $\rho_i$.  Note that $\Pr (m\!=\!+1|\sigma_l, \rho_i)= 1 - \Pr (m\!=\!-1|\sigma_l, \rho_i)$ and $\Pr (m\!=\!+1|\sigma_l, \rho_i)= 1 - \Pr (m\!=\!+1|\sigma_l, I-\rho_i)$, where $I$ is the identity matrix acting on the Hilbert space of the particle.}  
\begin{center}
\begin{tabular}{rclrcl}
\hline
$\Pr (m\!=\!+1|\sigma_1, \rho_+)$ &$=$& $1 - D_1$  &  $\Pr (m\!=\!-1|\sigma_1, \rho_+)$ &$= $ & $  D_1$ \\
$\Pr (m\!=\!+1|\sigma_1, \rho_-)$ &$=$ &$D_1$      &  $\Pr (m\!=\!-1|\sigma_1, \rho_-)$ &$=$ & $ 1 - D_1$  \\[3pt]
$\Pr (m\!=\!+1|\sigma_1, \rho_0)$ &$=$ & $ D_{+0}$   &  $\Pr (m\!=\!-1|\sigma_1, \rho_0)$ &$=$ & $1 - D_{+0}$ \\
$\Pr (m\!=\!+1|\sigma_1, \rho_1)$ &$=$ & $ 1-D_{+0}$ &  $\Pr (m\!=\!-1|\sigma_1, \rho_1)$ &$=$ & $ D_{+0}$\\[3pt]
$\Pr (m\!=\!+1|\sigma_3, \rho_+)$ &$= $ &$ D_{0+}$   &  $\Pr (m\!=\!-1|\sigma_3, \rho_+)$ &$=$ & $1 - D_{0+}$ \\
$\Pr (m\!=\!+1|\sigma_3, \rho_-)$ &$=$ & $ 1-D_{0+}$ &  $\Pr (m\!=\!-1|\sigma_3, \rho_-)$ &$=$ & $D_{0+}$\\[3pt]
$\Pr (m\!=\!+1|\sigma_3, \rho_0)$ &$=$ & $ D_3$      &  $\Pr (m\!=\!-1|\sigma_3, \rho_0)$ &$=$ & $1 - D_3$ \\
$\Pr (m\!=\!+1|\sigma_3, \rho_1)$ &$=$ & $ 1-D_3$    &  $\Pr (m\!=\!-1|\sigma_3, \rho_1)$ &$=$ & $D_3$ \\
\hline
\end{tabular}
\end{center}
\end{table}

Using the assumptions $D_{+0} = D_{0+} = 1/2$ and $D_1  = D_3 = D$,  the probabilities $\Pr(\rho_i, \sigma_l, a|b)$ that are needed to calculate $I(B\!:\!C)$ the mutual information are either $1/16$, $D/8$, or $(1-D)/8$.  The events corresponding to $\Pr(\rho_i, \sigma_l, a|b) = 1/16$ occur when Bob and Alice do not have a matching basis. The eight different compound events that occur when Alice and Bob have a matching basis can be organized into two mutually exclusive sets.  We define the set $\mathcal{C}_{\mu}$ to consist of the following four compound events: 
\begin{align*}
(\rho_-, \sigma_1, a\!=\!0), & & (\rho_+, \sigma_1, a\!=\!1 ), && (\rho_1, \sigma_3, a\!=\!0), && (\rho_0, \sigma_3, a\!=\!1) \ ;
\end{align*}
we define the set $\mathcal{C}_{\nu}$ to consist of the following four compound events:
\begin{align*}
(\rho_-, \sigma_1, a\!=\!1), & & (\rho_+, \sigma_1, a\!=\!0), && (\rho_1, \sigma_3, a\!=\!1), && (\rho_0, \sigma_3, a\!=\!0) \ .
\end{align*}
If the compound event $c$ is a member of $\mathcal{C}_{\mu}$ then $\Pr(c|b\!=\!0) = D/8$ and $\Pr(c|b\!=\!1) = (1-D)/8$.  If the compound event $c'$ is a member of $\mathcal{C}_{\nu}$ then $\Pr(c'|b\!=\!0) = (1-D)/8$ and $\Pr(c'|b\!=\!1) = D/8$.  

We are now prepared to determine the probability $\Pr(\mathbf{c}|b)$ of a compound-event string $\mathbf{c}$ given a particular value of $b$.   A compound-event string $\mathbf{c}$ can be described by three integers: $k$ its length (corresponding to the number of bit-announcements that Alice made), $x \leq k$ the number of shots in which Bob and Alice have a matching basis, and $y \leq x$ the number of compound events that fall within the set $\mathcal{C}_{\mu}$.   For a given number $k$ of bit-announcements, there is probability $\frac{k!}{x! k-x!} (1/2)^k$ of there being $x$ shots that also have a matching basis.  For a given $x$ shots that occur out of the $k$ with matching basis, there is probability $\frac{x!}{y!x-y!}D^y(1-D)^{x-y}$ that $y$ of the $x$ shots will fall within set $\mathcal{C}_{\mu}$.  %It is straightforward to show that the probability $\Pr(\mathbf{c}(k,x,y)|b)$ that a compound-event string is described by the values $(k,x,y)$ for a given value of $b$ is
%\[
%\Pr(\mathbf{c}(k,x,y)|b\!=\!0) =  \frac{\Pr(k)}{32^k}  D^y (1-D)^{x-y} \ ,
%\]
%and 
%\[
%\Pr(\mathbf{c}(k,x,y)|b\!=\!1) =  \frac{\Pr(k)}{32^k} (1-D)^y D^{x-y} \ .
%\]
It is straightforward to show that
\begin{multline*}
I(B\!:\!C) = \frac{1}{2} \sum_{\textbf{c}} \Biggl[ \sum_{b=0}^1  \Pr(\mathbf{c}(k,x,y)|b) \log \Pr(\mathbf{c}(k,x,y)|b)  \\
- \biggl( \sum_{b=0}^1 \Pr(\mathbf{c}(k,x,y) |b ) \biggr) \log \biggl( \sum_{b=0}^1 \Pr(\mathbf{c}(k,x,y) |b ) \biggr) \Biggr] \ ,
\end{multline*}
can be written as
\begin{multline*}
I(B\!:\!C) = \\ \frac{1}{2}  \sum_{k=0}^N \Pr(k) \sum_{x=0}^k \frac{k!}{x!k-x!}  \Bigl( \frac{1}{2}\Bigr)^k \sum_{y=0}^x   \frac{x!}{y!x-y!}  \Bigl( D^y (1-D)^{x-y} \Bigr) \log \Bigl( D^y (1-D)^{x-y} \Bigr)  \\ 
+ \Bigl( D^{x-y} (1-D)^{y} \Bigr) \log \Bigl( D^{x-y} (1-D)^{y} \Bigr) \\ -    \bigl( D^y (1-D)^{x-y} +  D^{x-y} (1-D)^{y} \bigr)  \log   \bigl( D^y (1-D)^{x-y} +  D^{x-y} (1-D)^{y} \bigr)  \ ,
\end{multline*}
which only depends upon $D$, $p_a$ and $N$.  This never becomes one because $\Pr(k=0)$ is always finite.  That is, the probability that Alice does not make a bit-announcement after $N$ shots is never nonzero.  Therefore the message will be noisy.  It is a question of how much noise Bob and Alice can tolerate that goes into their choice of $p_a$ and $N$.  Assuming that $D$ is a fixed value which will be dictated by the apparatus that Alice and Bob are using, for every value of $I(B\!:\!C)<1$ there is a family of $p_a$, $N$ pairs that lead to that particular value of $I(B\!:\!C)$.  As $p_a$ decreases, the value of $N$ increases to establish the same value of mutual information $I(B\!:\!C)$.  While increasing $N$ means that the protocol will take longer to complete, the effect of decreasing $p_a$ while increasing $N$ also has the effect of drastically increasing the number of result-announcements that Alice is expected to make, each of which provides Bob with another piece of data regarding how the state of the particle is changing as it travels from him to Alice.   The choice of $p_a$ and $N$ is therefore a trade-off between increased speed of communication (smaller $N$) and more data about the quantum channel (smaller $p_a$) once the desired value of $I(B\!:\!C)$ has been chosen.  

In the next section we discuss the ability of an eavesdropper to also obtain information about the message bit $b$ from Alice's bit-announcements, and the ability of Bob to detect the presence of any such eavesdropping activities through the use of Alice's result-announcements.

\section{The effectiveness and the effects of eavesdropping}

The goal of this section is twofold:  First, to demonstrate what an eavesdropper can learn about the message bit $b$ while this protocol is being carried out by Bob and Alice.  Second, to show how Bob can place a bound on what an eavesdropper may have learned after the $N$ shots by using the data he received from Alice's result-announcements.  To begin, we shall examine what an eavesdropper can learn if she is only listening to Alice's announcements and not interfering with any of the particles as they travel from Bob to Alice.  %We will refer to an eavesdropper acting in this simplistic way as a \textit{passive} eavesdropper.

It is important to emphasize that the eavesdropper, named Eve, does not have access to Bob's knowledge of the particular state in which he prepares each particle.  Therefore, when we use the mutual information to quantify what Eve learns over the course of Alice's $k$ bit-announcements, we must introduce a new random variable $E$ that describes the compound events corresponding to Eve's data on shots when Alice makes a bit-announcement.  The mutual information can be written as
\[
I(B:E) =  \sum_{\textbf{e}} \Biggl( \frac{1}{2} \sum_{b=0}^1  \Pr(\mathbf{e}|b) \log \Pr(\mathbf{e}|b) - \Pr(\mathbf{e}) \log \Pr(\mathbf{e}) \Biggr) \ 
\]
where the sum over $\mathbf{e}$ indicates that the summation is made over all possible compound-event strings, and where $\Pr (\mathbf{e}|b)$ is the probability of the compound-event string $\mathbf{e}$ given a particular value of the message bit $b$.   

When Eve is only listening to Alice's announcements, on each shot corresponding to a bit-announcement she records the compound event  $e = ( \sigma_l, a)$ which consists of the measurement type $\sigma_l$ and the bit-announcement $a$.  There are four possible compound events on each shot.  Each of these four events is equally likely from Eve's point of view.   This last statement depends upon Eve's calculation of $\Pr(m|\sigma_l)$, which is the probability that Alice has found the outcome $m$ given that her measurement corresponded to the operator $\sigma_l$.  Since Eve does not know the state in which Bob prepared the particle, it is equally likely to be in each of the four possible initial states.  Eve must calculate $\Pr(m|\sigma_l)$ using $\Pr(m|\sigma_l) = \frac{1}{4}\bigl(\Pr(m|\sigma_l, \rho_0) + \Pr(m|\sigma_l, \rho_1) +  \Pr(m|\sigma_l, \rho_+) + \Pr(m|\sigma_l, \rho_-) \bigr)  $, where $\Pr(m|\sigma_l, \rho_i)$ is the probability that Alice's measurement result is $m$ given that her measurement corresponds to $\sigma_l$ and that Bob prepared the particle in the state $\rho_i$.   This results in $\Pr(\sigma_l, a) = 1/4$ for each of the four possible compound events.  The random variable $E$ describes the strings $\mathbf{e} = (e_1, e_2, \ldots, e_k)$ of $k$ compound events for Eve, where $e_j$ is the compound event occurring on the $j$th shot that Alice has made a bit-announcement and where $0 \leq k \leq N$ is the number of bit-announcements made by Alice.  The probability $\Pr(\mathbf{e})$ of the string $\mathbf{e} = (e_1, e_2, \ldots, e_k)$ of compound events is the product of the probabilities for each individual compound event: $\Pr(\mathbf{e}) = \prod_{j=1}^k \Pr(e_j)$ and $\Pr(\mathbf{e}|b) = \prod_{j=1}^k \Pr(e_j|b)$.  When determining the value of $\Pr(e|b)$,  once again Eve must account for the fact that she does not know the state in which Bob prepared the particle.  After taking the weighted sum over all possible initial states  it is straightforward to show that $\Pr(e|b)$ is independent of the value of $b$ for each of the four possible compound events $e$.  Therefore, $\Pr(\mathbf{e}|b)$ is independent of the value of $b$ for every string of compound events $\mathbf{e}$.   From this, it is clear that $I(B:E)=0$.  Therefore, in order for Eve to learn anything about the message bit $b$ she must interact with the particles as they travel from Bob to Alice.

Eve has a choice of how she can interact with the particle.  She can apply a quantum operation to change the state of the particle as it travels from Bob to Alice, she can make a measurement (generally, a POVM) on the particle and then allow it to continue on its journey to Alice, or she can make the particle interact with an auxiliary ``probe'' system and wait for Alice to announce which measurement was made (corresponding to $\sigma_1$ or $\sigma_3$) before Eve makes a measurement of her own (generally, a POVM) on the probe system.  Any of these options, provided that it acts non-trivially, will change the correlations between the state in which Bob prepared the particle and the measurement result that Alice finds.  This change in the correlations will be made apparent to Bob through the result-announcements.  Therefore, it is Eve's goal to choose the interaction that improves her knowledge about the message bit $b$ the most (of any possible interaction) for a fixed disturbance to the statistics of Alice's measurement results.  

The lengthy calculation that determines which interaction accomplishes this goal for Eve is demonstrated in the Appendix.  It is based upon earlier work by Fuchs, Gisin, Niu, and Peres\cite{Fuchs_etal_1997} and utilizes an interaction with a probe system followed by a measurement.   Here we simply describe this interaction and evaluate its trade-off between Eve's expected gain in knowledge, measured by the mutual information, and the disturbance she causes, measured by the probability of a mismatch.  

The eavesdropping technique that optimizes what Eve learns for a fixed amount of disturbance is the following:  Eve prepares a probe system in some initial state $\ket{\psi}$.  As the particle is traveling from Bob to Alice, Eve ensures that the particle interacts with her probe system.  This interaction is described by the unitary operator $U$.  If Eve initially describes the state of the particle as $\rho_i$, then after this interaction the state of the combined system, particle plus probe, is described by her as $U (\rho_i \otimes \ket{\psi}\bra{\psi})U^{\dagger}$.  (This interaction generally forms an entangled state between the probe system and the particle.)  Eve then allows the particle to continue on its path to reach Alice, where it will be measured.   After Eve learns which measurement is used by Alice, Eve performs one of two measurements of her own on the probe system, depending upon which measurement Alice performed.

In order that she does not induce any noise that could be considered by Bob and Alice to be caused by some systematic error, Eve sets up her probe system $\ket{\psi}$ and interaction $U$ so that each of the four possible initial states $\rho_i$ (with $i=0,1,+,-$) evolves into the state $\rho_i' = (1-2d) \rho_i + d I$, where $I$ is the identity operator for the Hilbert space describing the particle's subsystem.   That is, if Eve knew that the initial state of the particle is $\rho_i$, then she would describe the state of the particle, as it reached Alice, as $\rho_i'=\Tr_2 \bigl(U (\rho \otimes \ket{\psi}\bra{\psi})U^{\dagger}\bigr)$, where $Tr_2$ indicates that the partial trace is taken over the degrees of freedom of the probe subsystem.   It is straightforward to show that when the state of the particle is described as $\rho_i' = (1-2d) \rho_i + d I$ and when Bob and Alice have a matching basis then the probability of Alice's measurement to result in a mismatch is $d$.  In this way, Eve chooses $\ket{\psi}$ and $U$ so that the parameter $d$ characterizes the amount of ``noise'' she introduces into the system.

This type of interaction is effective because the state of the combined particle plus probe system will change when Alice makes a measurement.  That is to say that if Eve knows that Bob prepares the particle in the state $\rho_i$ and that Alice's measurement outcome corresponds to the operator $\ket{f}\bra{f}$ (where $f=0,1,+,-$), then Eve describes the state of the probe system after the measurement as
\[
\Tr_1 \Bigl( \bigl( \ket{f}\bra{f} \otimes I_p\bigr)U \bigl(\rho_i \otimes \ket{\psi} \bra{\psi}\bigr)U^\dagger \bigl( \ket{f}\bra{f}  \otimes I_p\bigr)\Bigr) \ ,
\]
where $\Tr_1$ indicates taking the partial trace over the degrees of freedom of the particle subsystem, and where $I_p$ is the identity operator acting on the Hilbert space of the probe system.   If we then specify a particular measurement that Eve performs  on the probe system (by explicitly stating the POVM measurement operators), then the probabilities of Eve's possible measurement outcomes are easily calculable.  

As we described earlier, the optimal eavesdropping technique requires Eve to make two different measurements.  If Alice makes the measurement corresponding to $\sigma_1$ then Eve makes the measurement $\{ E_1, E_2, E_3, E_4 \}$, which has four outcomes that we describe by using its POVM operators.   If Alice makes the measurement corresponding to $\sigma_3$ then Eve makes the measurement $\{ F_1, F_2, F_3, F_4 \}$.  The specific form of $\ket{\psi}$, $U$, and the POVM operators $ E_1, \ldots E_4$, and $ F_1, \ldots, F_4$ are found in the Appendix.  

We use the mutual information between random variables $B$ and $E[d]$ to describe what Eve learns by employing this eavesdropping technique, where $d$ parameterizes the strength of Eve's interaction and $0 \leq d \leq 1/2$. We need to include the possible outcomes of Eve's measurement within the compound events that are described in the random variable $E[d]$.  In this way, on each shot the compound event $e = (\sigma_l, a, \mu)$ consists of Eve's measurement result $\mu$, as well as $\sigma_l$ the type of measurement announced by Alice, and $a$ the bit-announcement.   Eve's measurement result $\mu$ will be $E_1$, $E_2$, $E_3$, or $E_4$  when Alice announces $\sigma_1$ or $F_1$, $F_2$, $F_3$, or $F_4$ when Alice announces $\sigma_3$.  There are sixteen possible compound events which can occur for Eve on every shot corresponding to a bit-announcement.   The probability for each string $\mathbf{e}=(e_1, e_2, \ldots, e_k)$ is the product of the probabilities from each of the shots that compose that string  --- that is, $\Pr(\mathbf{e}) = \prod_{j=1}^k \Pr(e_j)$ and $\Pr(\mathbf{e}|b) = \prod_{j=1}^k \Pr(e_j|b)$.   

These sixteen events can be grouped into four different sets, with each compound event in the set occurring with the same probability given each of the two possible values of the message bit $b$.  Let $S_1$ indicate the set of all the compound events on the top line of Table \ref{events}, $S_2$ to indicate the set of all the events on the second line of Table \ref{events}, and so forth.  When the event $e_\alpha$ is in the set $S_1$ and the event $e_\beta$ is in the set $S_2$ then 
\begin{align*}
\Pr(e_\alpha | b\!=\!0) = \Pr(e_\beta|b\!=\!1)  = \frac{(1-d)}{2}\Bigl(1-2\sqrt{d(1-d)}\Bigr) &\equiv p_1 \ ,\\ 
\Pr(e_\alpha | b\!=\!1) = \Pr(e_\beta|b\!=\!0)  = \frac{(1-d)}{2}\Bigl(1+2\sqrt{d(1-d)}\Bigr) &\equiv p_2 \ .
\end{align*}
When the event $e_\gamma$ is in the set $S_3$ and the event $e_\delta$ is in the set $S_4$ then 
\begin{align*}
\Pr(e_\gamma | b\!=\!0) = \Pr(e_\delta|b\!=\!1)  = \frac{d}{2}\Bigl(1-2\sqrt{d(1-d)}\Bigr) &\equiv p_3 \ , \\
\Pr(e_\gamma | b\!=\!1) = \Pr(e_\delta|b\!=\!0)  = \frac{d}{2}\Bigl(1+2\sqrt{d(1-d)}\Bigr) &\equiv p_2 \ .
\end{align*}
The derivation of these probabilities can be found in the Appendix.
 
\begin{table} 
\caption{\label{events}The sixteen compound events which correspond to Alice's measurement type, Alice's bit-announcement, and a measurement result by Eve.  For events on the same line, the probability for each event to occur, for a given value of $b$, is the same.}  
\begin{center}
\begin{tabular}{rcccc}
\hline
$S_1$: $(\sigma_1, a\!=\!0, E_1)$, & $(\sigma_1, a\!=\!1, E_2)$, 
	& $(\sigma_3, a\!=\!0, F_1)$, & $(\sigma_3, a\!=\!1, F_2)$ \\
$S_2$: $(\sigma_1, a\!=\!0, E_2)$, & $(\sigma_1, a\!=\!1, E_1)$, 
	& $(\sigma_3, a\!=\!0, F_2)$, & $(\sigma_3, a\!=\!1, F_1)$ \\
$S_3$: $(\sigma_1, a\!=\!0, E_3)$, & $(\sigma_1, a\!=\!1, E_4)$, 
	& $(\sigma_3, a\!=\!0, F_3)$, & $(\sigma_3, a\!=\!1, F_4)$ \\
$S_4$: $(\sigma_1, a\!=\!0, E_4)$, & $(\sigma_1, a\!=\!1, E_3)$, 
	& $(\sigma_3, a\!=\!0, F_4)$, & $(\sigma_3, a\!=\!1, F_3)$ \\
\hline
\end{tabular}
\end{center}
\end{table}

When determining the probability of a string of events $\mathbf{e}$, it is sufficient to determine how many events occur in each of these categories $S_1,\ldots, S_4$.  If the string $\mathbf{e}(k_1, k_2, k_3, k_4)$ contains $k_i$ events from each set $S_i$, then the probability of that string of events to occur, given the two possible values of $b$, is
\begin{eqnarray*}
\Pr(\mathbf{e}{(k_1, k_2, k_3, k_4)}| b\!=\!0) &=& p_1^{k_1} p_2^{k_2} p_3^{k_3} p_4^{k_4} \\
\Pr(\mathbf{e}{(k_1, k_2, k_3, k_4)}| b\!=\!1) &=& p_1^{k_2} p_2^{k_1} p_3^{k_4} p_4^{k_3}\ . \\
\end{eqnarray*}
Given Eve's prior uncertainty about the value of $b$, the probability for a given string $\mathbf{e}{(k_1, k_2, k_3, k_4)}$ is found to be 
\[
\Pr(\mathbf{e}{(k_1, k_2, k_3, k_4)})= \frac{1}{2}\Bigl(p_1^{k_1} p_2^{k_2} p_3^{k_3} p_4^{k_4} + p_1^{k_2} p_2^{k_1} p_3^{k_4} p_4^{k_3} \Bigr) \ .
\] 

\begin{figure} [htbp]
%\vspace*{13pt}
\centerline{\epsfig{file=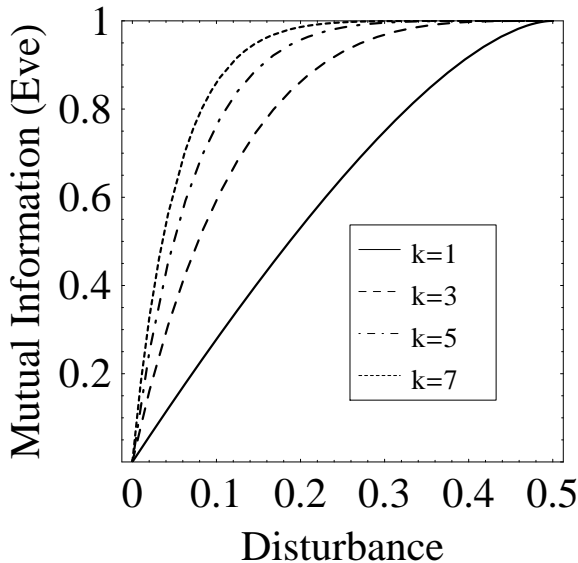, width=8.2cm}} %100 percent
\vspace*{13pt}
\caption{\label{infoplots}Mutual information as a function of distubance $D$, describing the amount an eavesdropper learns about the message bit given that Alice made $k$ bit-announcements.}
\end{figure}

We can now calculate the mutual information
\[
I(B\!:\!E[d]) =  \sum_{\textbf{e}} \Biggl( \frac{1}{2} \sum_{b=0}^1  \Pr(\mathbf{e}|b) \log \Pr(\mathbf{e}|b) - \Pr(\mathbf{e}) \log \Pr(\mathbf{e}) \Biggr) \ .
\]
Let us define the mutual information $I_k(B\!:\!E[d])$ to correspond to the situation when Alice makes exactly $k$ bit-announcements.  This can be written as
\begin{align*}
I_k (B\!:\!E[d]) &= \sum_{k_3 = 0}^{k-k_1-k_2} \sum_{k_2 = 0}^{k-k_1} \sum_{k_1 = 0}^k  
\frac{k!}{k_1!k_2!k_3!k_4!} \nonumber \\
&  \qquad \biggl( \half \Bigl[ \Pr\bigl(\mathbf{e}{(k_1, k_2, k_3, k_4)}| b\!=\!0\bigr) \log \Pr\bigl(\mathbf{e}{(k_1, k_2, k_3, k_4)}| b\!=\!0\bigr)  \nonumber  \\ 
& \qquad \qquad  +  \Pr\bigl(\mathbf{e}{(k_1, k_2, k_3, k_4)}| b\!=\!1\bigr) \log \Pr\bigl(\mathbf{e}{(k_1, k_2, k_3, k_4)}| b\!=\!1\bigr) \Bigr] \nonumber \\
& \qquad \qquad -      \Pr\bigl(\mathbf{e}{(k_1, k_2, k_3, k_4)}\bigr) \log \Pr\bigl(\mathbf{e}{(k_1, k_2, k_3, k_4)}\bigr) \biggr) \ .
\end{align*}
where $k_4 = k - k_1 - k_2 - k_3$.  The factor of 
\[
\frac{k!}{k_1!k_2!k_3!k_4!}
\]
is the number of different strings that share the same values of the $k_i$'s.  Due to the probabilistic nature of this process, Eve does not know the number of bit-announcements that will be made.  We have plotted in Figure \ref{infoplots} the mutual information versus disturbance when $k$ is $1$, $3$, $5$, or $7$.  It is easily seen that making more bit-announcements allows Eve to learn more for a fixed amount of disturbance, which is characterized by the parameter $d$.  The mutual information that is expected, given Eve's prior uncertainty for the number of bit-announcements $k$, can be taken account of by the weighted sum
\[
I(B\!:\!E[d])= \sum_{k=0}^N \Pr(k) I_k(B\!:\!E[d]) \ .
\]
This completes the analysis for determining what Eve expects to learn and the amount of noise she induces, measured by the probability of a mismatch, by employing the optimal eavesdropping techniques.   Next, we will show how Bob can place a bound on the amount of information an eavesdropper can be expected to know based upon Alice's result-announcements.

%%%%%%%
%old version
%
%\begin{figure}[hb]
%\begin{center}
%\includegraphics[width=18pc]{ff_edit_01.eps}
%\fcaption{\label{infoplots}Mutual information as a function of distubance $D$, describing the amount an eavesdropper learns about the message bit given that Alice made $k$ bit-announcements.}
%\end{center}
%\end{figure}
%
%%%%%%%%%%

Within the quantum cryptography literature, it is customary to attribute any noise to an eavesdropper's activities.  In keeping with this tradition, we will assume that any mismatches are due to Eve's activity, characterized by $d$, and not due to any mismatches that may be due to the apparatus that Alice and Bob are using, which we characterized by the value of $D$ earlier.  

During the transmission of the message, Bob will have a string of data describing the initial state in which he prepared $N$ particles.  He will also have the data from Alice's $k$ bit-announcements and from Alice's $N-k$ result-announcements.  If he detects $r$ mismatches ($0 \leq r \leq N-k$) during the result-announcements, then the experimental mean value of $d$ is $\overline d = r/(N-k)$.  If the statistics that he collects are sufficient so that the central limit theorem applies, then Bob can calculate the standard deviation $\sigma = \sqrt{r(N-k-r)/(N-k)}$.  Then he can state that with about a $95\%$ probability that any eavesdroppers present had obtained an equal or lesser knowledge of the message bit than any eavesdroppers using the ideal eavesdropping technique with a value of $d$ that was less than $d_{2 \sigma} = \overline d + 2 \sigma = d/(N-k) + 2\sqrt{d(N-k-d)/(N-k)}$.  And for every value of $d$, he can calculate $I_k (B\!:\!E[d])$.  So he knows, with a $2 \sigma$ confidence level, that any eavesdroppers know less than $I_k (B:E[d_{2 \sigma}])$ about the message bit.  If more confidence is necessary, Bob can simply go out to the appropriate number of standard deviations and make stronger statements.

Alternatively, a Bayesian statistical analysis could be used to employ all the data that Bob has available to him.   Broadly speaking, this would be used to calculate 
\[
\Pr(E[d]|\text{data}) = \frac{\Pr(\text{data}|E[d]) \Pr(E[d])}{\Pr(\text{data})},
\]
the probability that an eavesdropper is actively working at the level $d$ given all the data at hand, for every $d$.  Then, for example, Bob could determine the value of $\hat d$ such that $\int_{0}^{\hat d}\Pr(E[d]|\text{data}) \mathrm{d}d = 0.95$ in order to say that there is a probability of $95\%$ that any eavesdropper's knowledge of $b$ is less than $I_k(B:E[\hat d])$. (This is just an example of one type of statement that Bob could make when he knows $\Pr(E[d]|\text{data})$ for every $0 \leq d \leq 1/2$.)  An example of a piece of data that is not included in the simpler calculation is the probability of a mismatch caused by Bob's state preparation device and Alice's measurement device, which are assumed to be outside the influence of any eavesdroppers.  Another benefit of using this type of analysis would be its application to situations where there is not a sufficient amount of data for the central limit theorem to apply.  While we feel that this type of analysis will ultimately prove to be more effective than the simpler (but more established) techniques discussed earlier, we also feel that a detailed technical discussion of this type of Bayesian analysis would be more appropriate elsewhere.

\section{Discussion}
There are great similarities between the protocol introduced here, whose goal is to transfer a message to a privileged receiver while giving that receiver the opportunity to determine if that message has been exposed to any eavesdroppers during its transmission, and the BB84 quantum key distribution (QKD) protocol whose goal is to generate a private string of bits which is shared between two parties.  Indeed, the state preparation and measurements, along with their probabilities, are identical.  The two protocols deviate when it comes to the public announcements made by Alice.  (Actually, most descriptions of the BB84 protocol have Alice preparing the particles and sending them to Bob to be measured.  Here we have done the opposite to ensure that the message goes from Alice to Bob, which is common practice in the cryptography literature.  Since BB84 can be accomplished equally well either way, we have reversed the roles of Alice and Bob for our description of BB84.)   The public announcements phase of the BB84 protocol has three steps: Alice announces her measurement type (either $\sigma_1$ or $\sigma_3$), a number of announcements are made by both Alice and Bob to correct for errors, and then a number of announcements are made by both Alice and Bob to accomplish privacy amplification which makes their shared string of bits shorter but ensures that no eavesdroppers will know more than a negligible amount about their new shorter shared string of bits.    These last two steps, which require a great deal of public communication, are replaced by the bit-announcements and result-announcements by Alice.  

The private string of zeroes and ones that results from any QKD protocol is commonly referred to as a key, which can then be used for cryptographic tasks such as communicating secrets over a public communication channel or message authentication.  We refer to this type of key, the string of zeroes and ones, as a \textit{concrete key}.  However, we shall take a more general view of keys.  Following Shannon, we abstractly refer to a key as a mapping from the set of possible messages to the set of possible encrypted messages (cryptograms).  For the protocol introduced here, we can view the two possible values of Alice's message bit as the set of possible messages and her many possible public announcements as the set of possible cryptograms.  Furthermore, we can view the string of initial states that Bob prepares as his key.  Bob utilizes his knowledge of this string to decode the cryptogram and determine the value of the message bit.  Where this process differs greatly than those considered by Shannon\cite{Shannon45} is that the mapping that Bob's uses to determine the value of the message bit is probabilistic rather than deterministic.  Each of the two values of the message bit will have a non-zero probability for any given string of particles that Bob prepared and any given string of announcements by Alice, provided that there is noise in the system. 

When viewed in this way, by use of this more abstract notion of a key, the protocol introduced here can be thought of as the generation and use of a key --- on shots where bit-announcements occur --- while simultaneously checking the channel for noise which would be the hallmark of any eavesdropping activity.  Any protocol that includes these two factors would accomplish the same goal of being able to send a message while determining if that message had been exposed to any eavesdroppers.  What is accomplished is similar to the goals of tamper-indicating seals\cite{Johnston_06} which are attached to physical packages.  These tamper-indicating seals do not prevent anyone from opening the package, but they indicate whether or not they have been opened previously.  However, an advantage of the tamper-indicating seals that is not present with the protocol introduced here is that an effective tamper-indicating seal can be utilized by anyone who is opening the package, while here there is only a single privileged receiver who can check for eavesdroppers.  (To avoid confusion, it should be noted that there is another quantum cryptographic task that can be found in the literature under the name of a \textit{quantum seal}.\cite{Bechmann-Pasquinucci03, Bechmann-Pasquinucci05, Singh05, He05, Chau06}  The task introduced here is quite distinct from the quantum seal protocols that have been introduced, both in its goal and in its execution.  See \cite{Lopata_Bahder_07} for further comments.)

There have been a number of papers written on the topic of quantum secure direct communication (QSDC)\cite{Bostrom_etal_02, Wojcik_03, Deng_etal_03, Deng_etal_04, Lucamarini_etal_05}.  A careful reading of the literature seems to indicate that various authors who have written papers on this subject have assigned different meaning to the use of this term. (To exacerbate the issue, there are similar references that refer to an activity described as  quantum direct communication(QDC)\cite{Lee_etal_06} which seems to fall within the same general category as these others.)  It is worthwhile commenting on the similarity of the protocol introduced here to all those protocols introduced under the heading of QSDC (or QDC) because of their  similarity, both in the sense that they claim to be a quantum cryptographic task which differs from QKD and that they purport to transmit a message from one person to another.  The great difference arises from the fact that anyone who uses the protocol introduced here is willingly allowing an eavesdropper to intercept and read the transmitted message, at the peril of her eavesdropping activity being found out.  This is quite different from the cryptographic goals described in the QSDC literature -- most of the QSDC protocols purport to allow a message to be sent without the use of a private key shared between the two communicators and without any risk of an eavesdropper learning anything about the message.  

A possible advantage that may result in the use of the protocol introduced here is the reuse of concrete keys.  Say that a concrete key is privately shared between Alice and Bob.  Alice can encode a message using the concrete key and use the protocol introduced here to communicate that encrypted message to Bob.  When Bob receives the encrypted message he can decode it using their concrete key, but he can also determine if there have been any eavesdroppers trying to determine the encrypted message.  If he knows, to a high confidence level, that there have been no eavesdroppers active during Alice's transmission then he and Alice can have confidence to reuse their concrete key.  Or perhaps they may utilize a privacy amplification scheme to ensure that a smaller concrete key can be used to send future secret messages.  While there has not yet been any mathematical confirmation that this type of activity would be effective, we feel that it is worth careful investigation.  

A possible improvement of the protocol introduced here would be to include the use of an error correcting code by Bob and Alice to ensure that their noisy messages are transmitted accurately.  However, a new type of analysis must be undertaken, similar to the way that Slutsky \textit{et al}\cite{Slutsky_etal_1998} examined a more general class of eavesdropping activities than Fuchs \textit{et al}\cite{Fuchs_etal_1997} when they examined how the error-correcting stage of the  BB84 protocol effected an eavesdropper's advantages.

In summary, we have introduced a quantum cryptographic protocol that makes it possible to send a message to a privileged receiver so that receiver can detect the extent of which any eavesdroppers had access to that message during the course of its transmission.  We proved its effectiveness against all eavesdropping activities and discussed its relationship to other cryptographic tasks.  

%\begin{acknowledgments}
\section*{acknowledgements}
This work was funded in part by the Intelligence Advanced Research Projects Activity (IARPA) and by the Army Research Office (ARO).  This research was performed while P.A.L.\ held a National Research Council Research Associateship Award at the U.S.\ Army Research Laboratory.  He thanks T.\ Imbo and R.\ Espinoza for insightful discussions. 
%\end{acknowledgments}

%\section*{References}
\noindent
\bibliographystyle{unsrt}

\begin{thebibliography}{00}

\bibitem{Fuchs_etal_1997}
Christopher~A. Fuchs, Nicolas Gisin, Robert~G. Griffiths, Chi-Sheng Niu, and
  Asher Peres.
\newblock Optimal eavesdropping in quantum cryptography. i. information bound
  and optimal strategy.
\newblock {\em Physical Review A}, 56:1163--1172, 1997.

\bibitem{Shannon45}
Claude~E.\ Shannon.
\newblock Communication theory of secrecy systems.
\newblock In {\em Claude Elwood Shannon Collected Papers}, pages 84--143.
  {IEEE} Press, 1993.

\bibitem{Johnston_06}
Roger Johnston.
\newblock Tamper-indicating seals.
\newblock {\em American Scientist}, 94:515, 2006.

\bibitem{Bechmann-Pasquinucci03}
H.\ Bechmann-Pasquinucci.
\newblock Quantum seals.
\newblock {\em International Journal of Quantum Information}, 1:217, 2003.

\bibitem{Bechmann-Pasquinucci05}
H.~Bechmann-Pasquinucci, G.M. D'Ariano, and C.~Macchiavello.
\newblock Impossibility of perfect sealing of classical information.
\newblock {\em International Journal of Quantum Information}, 3:435, 2005.

\bibitem{Singh05}
S.K. Singh and R.~Srikanth.
\newblock Quantum seals.
\newblock {\em Physica Scripta}, 71:433, 2005.

\bibitem{He05}
Guang-Ping He.
\newblock Upper bounds of a class of imperfect quantum sealing protocols.
\newblock {\em Phys. Rev. A}, 71:054304, 2005.

\bibitem{Chau06}
H.M. Chau.
\newblock Insecurity of imperfect quantum bit seal.
\newblock {\em Physics Letters A}, 354:31, 2006.

\bibitem{Lopata_Bahder_07}
Paul~A. Lopata and Thomas~B. Bahder.
\newblock The effectiveness of quantum operations for eavesdropping on sealed
  messages.
\newblock {\em Journal of Physics:Conference Series}, 70:012011, 2007.

\bibitem{Bostrom_etal_02}
Kim {B\"ostrom} and Timo Felbinger.
\newblock Deterministic secure direct communication using entanglement.
\newblock {\em Physical Review Letters}, 89:187902, 2002.

\bibitem{Wojcik_03}
Antoni W\'ojcik.
\newblock Eavesdropping on the ``ping-pong'' quantum communication protocol.
\newblock {\em Physical Review Letters}, 90:157901, 2003.

\bibitem{Deng_etal_03}
Fu-Guo Deng, Gui~Lu Long, and Xiao-Shu Liu.
\newblock Two-step quantum direct communication protocol using the
  einstein-podolsky-rosen pair block.
\newblock {\em Physical Review A}, 68:042317, 2003.

\bibitem{Deng_etal_04}
Fu-Guo Deng and Gui~Lu Long.
\newblock Secure direct communication with a quantum one-time pad.
\newblock {\em Physical Review A}, 69:052319, 2004.

\bibitem{Lucamarini_etal_05}
Marco Lucamarini and Stefano Mancini.
\newblock Secure deterministic communication without entanglement.
\newblock {\em Physical Review Letters}, 94:140501, 2005.

\bibitem{Lee_etal_06}
Hwayean Lee, Jongin Lim, and HyungJin Yang.
\newblock Quantum direct communication with authentication.
\newblock {\em Physical Review A}, 73:042305, 2006.

\bibitem{Slutsky_etal_1998}
Boris~A. Slutsky, Ramesh Rao, Pang-Chen Sun, and Y.~Fainman.
\newblock Security of quantum cryptography against individual attacks.
\newblock {\em Physical Review A}, 57:2383, 1998.

\bibitem{Chefles_2000}
Anthony Chefles.
\newblock Quantum state discrimination.
\newblock {\em Contemporary Physics}, 41:401--424, 2000.

\end{thebibliography}

\appendix

\setcounter{section}{1}

\subsection{Plan for this section}
In this Appendix we determine the optimal eavesdropping technique for the eavesdropper, Eve.  This technique maximizes what she expects to learn, measured by the mutual information, for a given disturbance she causes, quantified by the probability of a mismatch.  (As was discussed earlier, a mismatch occurs on a shot when Bob and Alice have a matching basis, but where Alice's measurement result is opposite to the one that Bob expected based upon his knowledge of the state in which he prepared the particle.)  Our analysis relies heavily upon work done by Fuchs, Gisin, Griffiths, Niu, and Peres\cite{Fuchs_etal_1997} in which they studied certain classes of attacks on the BB84 protocol.  

Their analysis focuses on the best eavesdropping activity when the eavesdropper acts on each particle individually, as opposed to the eavesdropper using some ``large'' probe system that interacts with all the particles from $N$ shots before making a measurement of the probe.  Those that are studied by Fuchs \textit{et al} are referred to as \textit{individual attacks}, as opposed to more general types of attacks which are referred to as  \textit{collective attacks}\cite{Slutsky_etal_1998} and \textit{joint attacks}\cite{Slutsky_etal_1998}.  While later analyses\cite{Slutsky_etal_1998} of eavesdropping on BB84 considered these more general types of eavesdropping activities, this is not necessary in our case.  The reason for this is because in BB84 an eavesdropper can infer certain global properties of Alice and Bob's shared string of bits from the announcements that are made during the error correction phase of that protocol, but here there are no such error correction announcements.  Therefore, neither a collective attack nor a joint attack will provide any benefit to an eavesdropper because the protocol does not divulge any further data that an eavesdropper could use to inform her decision as to which specific activity, from within these two classes of the more general types of attacks, she would undertake.  Therefore, our analysis, which will parallel the analysis by Fuchs \textit{et al} and only focus on activities in which the eavesdropper interacts with each particle individually, encompasses the most general type of eavesdropping activities that we need to consider.

But we cannot simply apply the Fuchs \textit{et al} results directly to our situation because the details of BB84 are distinct from ours in one critical aspect.  In BB84, if on a particular shot Bob and Alice do not have a matching basis, then they will not make use of that measurement result as a bit for their key.  In the protocol introduced here, it is possible that Bob and Alice do not have a matching basis on a particular shot, but Eve can still use Alice's announcement from that shot to determine the message bit.  Fortunately, we can utilize many of their results to determine the eavesdropping activity that results in the most information gained by the eavesdropper for a fixed amount of disturbance.

We summarize the methods employed by Fuchs \textit{et al} because our present analysis depends upon a number of specific points within their analysis.  Instead of simply listing these specific points, we shall follow the logic used by Fuchs \textit{et al} to reach these points.  Furthermore, once these points are in hand, the analysis for our new protocol can be accomplished quickly and clearly.

\subsection{Summary of the Fuchs \textit{et al} optimization}\label{Fuchs}

The eavesdropping activities considered by Fuchs \textit{et al} are based upon the concept of an eavesdropping probe system which was discussed earlier.  As the particle, described by state $\rho_i$, travels from Bob to Alice, Eve ensures that it interacts with a probe system she had prepared in the state $\ket{\psi}$.    The state of the combined system evolves unitarily and then the original particle travels to Alice while the probe system stays under the control of Eve.  If Alice announces that her measurement corresponds to $\sigma_1$ then Eve will make a (POVM) measurement described by the operators $\{ E_1, E_2, E_3, E_4 \}$, and if Alice announces that her measurement corresponds to $\sigma_3$ then Eve will make a different (POVM) measurement described by the operators $\{ F_1, F_2, F_3, F_4 \}$.  

The analysis performed by Fuchs \textit{et al} was to find a bound on the the mutual information, describing what Eve expects to learn, for a fixed probability of a mismatch induced by her action.  We use the symbol $d_1$ ($d_3$) to indicate the probability of a mismatch if Alice's measurement corresponds to $\sigma_1$ ($\sigma_3$, respectively).    We use the symbol $\mathcal{I}_1$ ($\mathcal{I}_3$) to describe the what Eve learns --- quantified using the mutual information between the random variable describing both possible bit values and the random variable describing all  outcomes of possible ``probe and measurement'' procedures --- when Bob and Alice have a matching basis.  (Recall that in the BB84 protocol, Alice and Bob only use the data resulting from shots when they have a matching basis.) Fuchs \textit{et al} showed, by the application of a number of inequalities, that the relations 
\begin{eqnarray} \label{xinfozdisturbance}
\mathcal{I}_1 & \leq& \half \biggl( \Bigl[1+2\sqrt{d_3 (1+d_3)}\Bigr] \log \Bigl[1+2\sqrt{d_3(1+d_3)}\Bigr] \nonumber \\ 
&& \qquad \qquad + \Bigl[1-2\sqrt{d_3(1+d_3)}\Bigr] \log \Bigl[1-2\sqrt{d_3(1+d_3)}\Bigr]   \biggr) \ , \\
\label{zinfoxdisturbance}
\mathcal{I}_3 & \leq& \half \biggl( \Bigl[1+2\sqrt{d_1 (1+d_1)}\Bigr] \log \Bigl[1+2\sqrt{d_1(1+d_1)}\Bigr]  \nonumber \\
 && \qquad \qquad + \Bigl[1-2\sqrt{d_1(1+d_1)}\Bigr] \log \Bigl[1-2\sqrt{d_1(1+d_1)}\Bigr]   \biggr) \ 
\end{eqnarray}
describe the best an eavesdropper can do in terms of a trade-off between information learned and disturbance caused.  They then show that their proof allows for an eavesdropping activity that maximizes both of these inequalities simultaneously (so that the $\leq$ is replaced by $=$). 

For the second part of their analysis, they demonstrate a particular type of probe-and-measurement activity that saturates these bounds just discussed.  They call this probe ``symmetric'' because, among other things, $\mathcal{I}_1 = \mathcal{I}_3$ and $d_1 = d_3$.  (We shall discuss the other criteria that make it ``symmetric'' shortly.)  Within the BB84 protocol, half the time Alice's measurement corresponds to $\sigma_1$ and half the time it corresponds to $\sigma_3$ (while ``throwing out'' the shots where Alice and Bob do not have a matching basis),  so the expected mutual information is $\mathcal{I} = \half(\mathcal{I}_1 + \mathcal{I}_3)$ and the expected probability of a mismatch is $d = \half(d_1 + d_3)$.  In this case, Equation \ref{xinfozdisturbance} can be rewritten in terms of $\mathcal{I}$ and $d$:
\begin{eqnarray}\label{info_disturbance}
\mathcal{I} & \leq& \half \biggl( [1+2\sqrt{d (1+d)}] \log [1+2\sqrt{d(1+d)}] \nonumber \\ 
& & \qquad \qquad + [1-2\sqrt{d(1+d)}] \log [1-2\sqrt{d(1+d)}]   \biggr) \ .
\end{eqnarray}

The ansatz outlined by Fuchs \textit{et al} works as follows: First, the particle, described by state $\ket{i}$ (with $i = 0,1,+,-$), interacts with a probe system prepared in some initial state $\ket{\psi}$.  The combined system, described by the state $\ket{i} \otimes \ket{\psi}$, evolves unitarily in the following way:
\begin{eqnarray*}
\ket{0} \otimes \ket{\psi} & \stackrel{U}{\longrightarrow} & \sqrt{1-d} \ket{0} \otimes \ket{\chi_{00}}  + \sqrt{d} \ket{1} \otimes \ket{\chi_{01}} \\
\ket{1} \otimes \ket{\psi} &\stackrel{U}{\longrightarrow} &  \sqrt{d} \ket{0} \otimes \ket{\chi_{10}} + \sqrt{1-d} \ket{1} \otimes \ket{\chi_{11}} \ .
\end{eqnarray*}
%(the notation we are using differs from that employed by Fuchs \textit{et al}). 
Such a unitary operator $U$ exists if and only if $\braket{\chi_{00}}{\chi_{10}}=0$ and $\braket{\chi_{01}}{\chi_{11}}=0$.  This type of interaction was chosen so that $\rho_i = \ket{i}\bra{i}$ (for $i =  0,1$) evolves into $\rho_i'=(1-2d) \rho_i + d I$ after we trace over the degrees of freedom of the probe system.  In order for this to occur, it also requires that $\braket{\chi_{00}}{\chi_{01}} = 0$ and $\braket{\chi_{10}}{\chi_{11}} = 0$. 

Fuchs \textit{et al} also desire for $U$ to act in a similar way on the initial states $\rho_{+}$ and $\rho_{-}$:
\begin{eqnarray*}
\ket{+} \otimes \ket{\psi} & \stackrel{U}{\longrightarrow}& \sqrt{1-d} \ket{+} \otimes \ket{\chi_{++}} + \sqrt{d} \ket{-} \otimes \ket{\chi_{+-}} \\
\ket{-} \otimes \ket{\psi} &\stackrel{U}{\longrightarrow}& \sqrt{d} \ket{+} \otimes \ket{\chi_{-+}} + \sqrt{1-d} \ket{-} \otimes \ket{\chi_{--}} \ .
\end{eqnarray*}
The unitarity of $U$ requires $\braket{\chi_{++}}{\chi_{-+}}=0$ and $\braket{\chi_{+-}}{\chi_{--}}=0$.   In order for $\rho_i$ to evolve into $\rho_i' = (1-2d) \rho_i + d I$, for $i=+,-$ (after we trace over the degrees of freedom of the probe system), it is necessary that $\braket{\chi_{++}}{\chi_{+-}} = 0$ and $\braket{\chi_{-+}}{\chi_{--}} = 0$.   The linear relationships between $\ket{0}, \ket{1}, \ket{+}$, and $\ket{-}$ allow us to write  
\begin{eqnarray*}
\ket{\chi_{++}}&=& \frac{1}{2}\Bigl[ \ket{\chi_{00}} + \ket{\chi_{11}} + \sqrt{\frac{d}{1-d}}\bigl(  \ket{\chi_{01}} + \ket{\chi_{10}}\bigr)   \Bigr]  \\
\ket{\chi_{+-}}&=& \frac{1}{2}\Bigl[ \sqrt{\frac{1-d}{d}} \bigl( \ket{\chi_{00}} - \ket{\chi_{11}} \bigr) +   \ket{\chi_{01}} - \ket{\chi_{10}}   \Bigr]  \\
\ket{\chi_{-+}}&=& \frac{1}{2}\Bigl[ \sqrt{\frac{1-d}{d}} \bigl( \ket{\chi_{00}} - \ket{\chi_{11}} \bigr) -   \bigl(\ket{\chi_{01}} - \ket{\chi_{10}}\bigr)   \Bigr]  \\
\ket{\chi_{--}}&=& \frac{1}{2}\Bigl[ \ket{\chi_{00}} + \ket{\chi_{11}} - \sqrt{\frac{d}{1-d}}\bigl(  \ket{\chi_{01}} + \ket{\chi_{10}}\bigr)   \Bigr] \label{linearrelationships} \ .
\end{eqnarray*}
From these linear relationships, and the earlier requirements on the orthogonalities between the various $\ket{\chi_{ij}}$ probe states, it is easy to show that $\braket{\chi_{00}}{\chi_{11}} = \braket{\chi_{11}}{\chi_{00}}$ and $\braket{\chi_{01}}{\chi_{10}} = \braket{\chi_{10}}{\chi_{01}}$, and that 
\begin{equation} \label{Dandchi}
d = \frac{1 - \braket{\chi_{00}}{\chi_{11}}}{1 - \braket{\chi_{00}}{\chi_{11}} + \braket{\chi_{01}}{\chi_{10}}} \ .
\end{equation}

Fuchs \textit{et al} argue that these restrictions on the evolution of the combined system (particle plus probe) we have just described are not a hindrance to their task of determining the optimal eavesdropping activities.  Their reasoning for this is nicely discussed in Appendix B of their manuscript, but can be summarized by saying that for every ``non-symmetric'' activity --- one for which the requirements that (i) $d_1 = d_3$, (ii) $\mathcal{I}_3 = \mathcal{I}_1$, and (iii) $\rho_i$ evolves into $(1-2d) \rho_i + d I$ for all $i = 0,1,+,-$, are not all satisfied --- there exists some symmetric activity (where requirements (i)--(iii) \textit{are} satisfied) that achieves an equal amount of mutual information while causing the same probability for a mismatch.

Once the symmetric evolution has been described, and it has been established that the optimal eavesdropping activity can fall within this form, Fuchs \textit{et al} go on to explain how this type of evolution can lead to the eavesdropper determining the value of the key bit.  In their BB84 analysis, the key bit has value $0$ if Bob prepares the particle in the state $\rho_0$ or $\rho_+$ and the key bit has value $1$ if Bob prepares the particle in the state $\rho_1$ or $\rho_-$.  

Let us assume, as Fuchs \textit{et al} have done, that Bob prepares the particle in the state $\rho_{0}$ and Alice makes the measurement corresponding to $\sigma_3$.  It is easy to see that Alice has probability $(1-d)$ to find the result $m=+1$ and probability $d$ to finds the result $m=-1$.  If Bob prepares the particle in the state $\rho_{0}$ and Alice finds the result $m=+1$ then the resulting state of the probe will be $\ket{\chi_{00}}$ after tracing over the degrees of freedom of the particle.  And if Alice finds the result $m=-1$ then the resulting state of the probe will be $\ket{\chi_{01}}$.   This means that Eve's best description of the state of the probe is 
\[
\Gamma_0 = (1-d)\ket{\chi_{00}}\bra{\chi_{00}} + d \ket{\chi_{01}} \bra{\chi_{01}}
\]
if she knows that Bob prepared the state $\ket{0}$ and Alice's measurement corresponded to $\sigma_3$.  By similar reasoning, if Eve knows that Bob prepared the state $\rho_{1}$ and Alice made the same type of measurement then Eve's best description of the state of the probe is 
\[
\Gamma_1 =  d \ket{\chi_{10}} \bra{\chi_{10}} + (1-d)\ket{\chi_{11}}\bra{\chi_{11}} \ .
\]  
The problem of determining the state in which Bob prepared the particle then becomes a problem of distinguishing between the two mixed states $\Gamma_0$ and $\Gamma_1$.  

In other words, on every shot in which Alice's measurement corresponds to $\sigma_3$ and she and Bob have a matching basis, there is a one-half probability that the state of the probe is $\Gamma_0$ and a one-half probability that the state of the probe is $\Gamma_1$.  If Eve can determine which state the probe is in then she knows which state Bob prepared.

A great deal of work has been done on distinguishing between quantum states,\cite{Chefles_2000} but how to optimally distinguish between two mixed states is still an open question.  However, Fuchs \textit{et al} have found a procedure which is optimal for distinguishing between the states $\Gamma_0$ and $\Gamma_1$ when there is an equal chance that the system is in either state.  The reason why we know that they have found the optimal procedure for distinguishing between $\Gamma_0$ and $\Gamma_1$ is the following: The mutual information that describes how much Eve learns about which state Bob prepared is equal to the mutual information that describes Eve's ability to distinguish $\Gamma_0$ from $\Gamma_1$.  The procedure they have found (which we will describe shortly) allows them to maximize the inequality which we have re-written in Equation (\ref{info_disturbance}).  If there were some other procedure that allowed Eve to better distinguish between $\Gamma_0$ and $\Gamma_1$, this would increase the mutual information without effecting the disturbance caused, which would violate the inequality in Equation (\ref{info_disturbance}).   Therefore, the procedure that Fuchs \textit{et al} found for distinguishing between $\Gamma_0$ and $\Gamma_1$, when both have equal probability, is optimal.  

We now shall explain the procedure that Fuchs \textit{et al} have devised to optimally distinguish between the equiprobable mixed states $\Gamma_0$ and $\Gamma_1$.   Our description of their procedure differs from theirs, but the procedure itself is equivalent.  

Let us define a four-dimensional orthonormal basis for the state of the probe particle, $ \{ \ket{\eta_0}, \ket{\eta_1}, \ket{\eta_2}, \ket{\eta_3} \}$ with $\braket{\eta_i}{\eta_j} = \delta_{ij}$.    (Where $\delta_{ij}=1$ when $i=j$ and $\delta_{ij} = 0$ when $i\neq j$.)  Take the $\ket{\chi_{ij}}$'s to be defined as follows: 
\begin{eqnarray*}
\ket{\chi_{00}}&=&\ket{\eta_0}\\
\ket{\chi_{01}}&=&\ket{\eta_1}\\
\ket{\chi_{10}}&=&(1-2d) \ket{\eta_1} + 2\sqrt{d(1-d)}\ket{\eta_3}\\
\ket{\chi_{11}}&=&(1-2d) \ket{\eta_0} + 2\sqrt{d(1-d)}\ket{\eta_2} \ .
\end{eqnarray*}
Furthermore, we introduce the angle $\gamma$ so that $\cos \gamma = ([1 + 2\sqrt{d(1-d)}]/2)^{1/2}$ and $\sin \gamma = ([1 - 2\sqrt{d(1-d)}]/2)^{1/2}$.  Eve performs a measurement (POVM) on the probe particle that is described by the operators $\{ E_1, E_2, E_3, E_4 \}$, where $E_i = \ket{\varepsilon_i}\bra{\varepsilon_i}$ for $i=1,2,3,4$, and where the $\ket{\varepsilon_i}$'s form a different orthonormal basis,
\begin{eqnarray*}
\ket{\varepsilon_1}&=&\cos \gamma \ket{\eta_1} - \sin \gamma \ket{\eta_3}\\
\ket{\varepsilon_2}&=&\sin \gamma \ket{\eta_1} + \cos \gamma \ket{\eta_3}\\
\ket{\varepsilon_3}&=&\cos \gamma \ket{\eta_2} - \sin \gamma \ket{\eta_4}\\
\ket{\varepsilon_4}&=&\sin \gamma \ket{\eta_2} + \cos \gamma \ket{\eta_4} \ .
\end{eqnarray*}

Recall that it is Eve's goal to determine whether the probe particle is in the state $\Gamma_0 = (1-d)\ket{\chi_{00}}\bra{\chi_{00}} + d \ket{\chi_{01}}\bra{\chi_{01}}$, which would indicate that Bob prepared his particle in the state $\ket{0}$, or whether the probe is in the state $\Gamma_1 = (1-d)\ket{\chi_{11}}\bra{\chi_{11}} + d \ket{\chi_{10}}\bra{\chi_{10}}$ which would indicate that Bob had prepared his particle in the state $\ket{1}$.  For this purpose, it is necessary to calculate $\Pr(\Gamma_j | E_i )$, which is the probability that the probe particle was in the state $\Gamma_j$ (with $j=0,1$) given that her measurement result corresponded to $E_i$ (with $i=1,2,3,4$).  Since both states, $\Gamma_0$ and $\Gamma_1$, were equiprobable before the measurement, a simple application of Bayes' rule shows that $\Pr(\Gamma_j | E_i )= \half \Pr(E_i| \Gamma_j)/\Pr(E_i)$, where $\Pr(E_i| \Gamma_j)$ is the probability of a particular measurement outcome for Eve, described by $E_i$ ($i = 1,2,3,4$), given that the state of the probe is $\Gamma_j$ ($j=0,1$), and $\Pr(E_i)=\half[\Pr(E_i| \Gamma_0)+\Pr(E_i| \Gamma_1)]$.  These probabilities are easily calculable and are summarized in Table \ref{distinguishing}.

\begin{table}
\caption{The probability of a particular measurement outcome for Eve, described by $E_i$, given that the state of the probe is $\Gamma_j$. Also, $\sin^2 \gamma = \half [1 - 2 \sqrt{d(1-d)}]$ and $\cos^2 \gamma = \half [1 + 2 \sqrt{d(1-d)}]$.}
\begin{center}
\begin{tabular}{rcc}
\label{distinguishing}
 & \multicolumn{2}{c}{$\Pr(E_i| \Gamma_j)$} \\ \cline{2-3}
 & $\Gamma_0$ & $\Gamma_1$ \\
\hline \hline
$E_1:$ &  $(1-d)\cos^2 \gamma$ & $(1-d)\sin^2 \gamma$ \\
$E_2:$ &  $(1-d)\sin^2 \gamma$ & $(1-d)\cos^2 \gamma$ \\
$E_3:$ &  $d\cos^2 \gamma$ & $d\sin^2 \gamma$ \\
$E_4:$ &  $d\sin^2 \gamma$ & $d\cos^2 \gamma$ \\
\hline \hline
\end{tabular}
\end{center}
\end{table}

From these results we can determine what Eve learns by employing this type of eavesdropping activity, which is quantified by the mutual information between the random variable describing the two possible states $\rho_{0}$ and $\rho_{1}$, each with probability $1/2$ and random variable describing the possible measurement outcomes by Eve, $\{ E_1, E_2, E_3, E_4 \}$.  It is found to be 
\begin{eqnarray*}
\mathcal{I}_3 & =& \half \biggl( \Bigl[1+2\sqrt{d (1+d)}\Bigr] \log \Bigl[1+2\sqrt{d(1+d)}\Bigr] \\ 
&& \qquad \qquad + \Bigl[1-2\sqrt{d(1+d)}\Bigr] \log \Bigl[1-2\sqrt{d(1+d)}\Bigr]   \biggr) \ ,
\end{eqnarray*}
which saturates the bound in Equation (\ref{zinfoxdisturbance}). (Recall that in this symmetric case $d_1 = d$.)  

The description of this procedure for distinguishing $\Gamma_0$ from $\Gamma_1$ was focused on the case when Alice announces that her measurement corresponded to $\sigma_3$.  An analogous procedure can be followed to distinguish the mixed state 
\[
\Gamma_+ = (1-d) \ket{\chi_{++}}\bra{\chi_{++}} + d \ket{\chi_{+-}}\bra{\chi_{+-}}
\]
from 
\[
\Gamma_- = d \ket{\chi_{-+}}\bra{\chi_{-+}} + (1-d) \ket{\chi_{--}}\bra{\chi_{--}} \   
\]
which are the resulting states of the probe when Alice's measurement corresponds to $\sigma_1$ and when she and Bob have a matching basis.  It can be shown that such a procedure results in    
\begin{eqnarray*}
\mathcal{I}_1 & =& \half \biggl( [1+2\sqrt{d (1+d)}] \log [1+2\sqrt{d(1+d)}] \\ 
&& \qquad \qquad + [1-2\sqrt{d(1+d)}] \log [1-2\sqrt{d(1+d)}]   \biggr) \ ,
\end{eqnarray*}
which saturates the bound in Equation (\ref{xinfozdisturbance}).  Because the two types of measurement that Alice performs occur with equal probability, it is easy to see that $\mathcal{I}=\half (\mathcal{I}_1 + \mathcal{I}_3)$ so that the inequality shown in Equation (\ref{info_disturbance}) is saturated.  This completes our description of the optimal method Fuchs \textit{et al} found to eavesdrop on the non-error-corrected string of key bits which result from the BB84 process.  We will rely upon this analysis and their results extensively to demonstrate the optimal bound for eavesdropping on the protocol introduced here. 

\subsection{Optimal eavesdropping activity for this current protocol}
Our discussion of the optimal activity of an eavesdropper shall begin with a short reminder of the goal of the eavesdropper.  Since, on every shot that a bit-announcement is made, there will be an announced bit $a$ which depends upon the measurement result, $m$, and the value of the message bit $b$, it is the eavesdropper's intention to infer the value of $m$.  This is a different question than the one which Fuchs \textit{et al} answered, which was to determine the state in which Bob prepared the particle.  

In order to determine which measurement result was found by Alice, on a particular shot, we will assume without loss of generality that Eve uses a symmetric probe for eavesdropping.  The unitary evolution of the coupled particle and probe systems states is characterized by 
\begin{eqnarray*}
\ket{0} \otimes \ket{\psi} & \stackrel{U}{\longrightarrow} & \sqrt{1-d} \ket{0} \otimes \ket{\chi_{00}} + \sqrt{d} \ket{1} \otimes \ket{\chi_{01}} \\
\ket{1} \otimes \ket{\psi} &\stackrel{U}{\longrightarrow}& \sqrt{d} \ket{0} \otimes \ket{\chi_{10}} + \sqrt{1-d} \ket{1} \otimes \ket{\chi_{11}} \\
\ket{+} \otimes \ket{\psi} & \stackrel{U}{\longrightarrow}& \sqrt{1-d} \ket{+} \otimes \ket{\chi_{++}} + \sqrt{d} \ket{-} \otimes \ket{\chi_{+-}} \\
\ket{-} \otimes \ket{\psi} &\stackrel{U}{\longrightarrow}& \sqrt{d} \ket{-} \otimes \ket{\chi_{-+}} + \sqrt{1-d} \ket{-} \otimes \ket{\chi_{--}} 
\end{eqnarray*}
subject to the conditions 
\begin{gather*}
\braket{\chi_{00}}{\chi_{10}} = \braket{\chi_{01}}{\chi_{11}} = \braket{\chi_{00}}{\chi_{01}} = \braket{\chi_{10}}{\chi_{11}} = 0 \ , \\
\braket{\chi_{++}}{\chi_{-+}} = \braket{\chi_{+-}}{\chi_{--}} = \braket{\chi_{++}}{\chi_{+-}} = \braket{\chi_{-+}}{\chi_{--}} = 0 \  , \\
\braket{\chi_{00}}{\chi_{11}} = \braket{\chi_{11}}{\chi_{00}}, \quad \braket{\chi_{01}}{\chi_{10}} = \braket{\chi_{10}}{\chi_{01}} \ , \\
\braket{\chi_{++}}{\chi_{--}} = \braket{\chi_{--}}{\chi_{++}}, \quad \braket{\chi_{+-}}{\chi_{-+}} = \braket{\chi_{-+}}{\chi_{+-}} \ , \\
d  = \frac{1 - \braket{\chi_{00}}{\chi_{11}}}{1 - \braket{\chi_{00}}{\chi_{11}} + \braket{\chi_{01}}{\chi_{10}}} \ .
\end{gather*}

When this type of evolution occurs, the state of the particle $\rho_i$ ($i=0,1,+,-$) evolves into $\rho_i'=(1-2d)\rho_i + dI$ with $0 \leq d \leq 1/2$.  The probability that there will be a mismatch  on a particular shot, given that Alice and Bob have a matching basis, is $d$.      

Let us now focus on the case when Alice makes the measurement corresponding to $\sigma_3$ and finds the result $m=+1$.  The state of the probe that results from Bob preparing the particle in each of the four possible initial states is shown below.  We use the notation $\Gamma_{(\sigma_l, m, \rho_{i})}$ to indicate the resulting state of the probe given that Alice's measurement corresponded to $\sigma_l$ (with $l = 1,3$) and that she found the result $m$ after Bob prepared the particle in the state $\rho_{i}$ (where $i=0,1,+,-$).  These states are the following
\begin{eqnarray*}
\Gamma_{(\sigma_3, m=+1, \rho_{0})} &=& \ket{\chi_{00}}\bra{\chi_{00}}\\
\Gamma_{(\sigma_3, m=+1, \rho_{1})} &=& \ket{\chi_{10}}\bra{\chi_{10}}\\
\Gamma_{(\sigma_3, m=+1, \rho_{+})} &=& (1-d) \ket{\chi_{00}}\bra{\chi_{00}} + d \ket{\chi_{10}}\bra{\chi_{10}} \\
 && \quad + \sqrt{d(1-d)}\bigl[ \ket{\chi_{00}}\bra{\chi_{10}} + \ket{\chi_{10}}\bra{\chi_{00}} \bigr] \\
\Gamma_{(\sigma_3, m=+1, \rho_{-})} &=& (1-d) \ket{\chi_{00}}\bra{\chi_{00}} + d \ket{\chi_{10}}\bra{\chi_{10}} \\
 && \quad - \sqrt{d(1-d)}\bigl[ \ket{\chi_{00}}\bra{\chi_{10}} + \ket{\chi_{10}}\bra{\chi_{00}} \bigr]  \ .
\end{eqnarray*}
The probability for the event $(\sigma_l, m, \rho_{i})$ to occur is easily calculated from the fact that Alice is equally likely to make either measurement, Bob is equally likely to prepare each of four initial states, and the unitary evolution $U$ is designed so that each initial state of the particle $\rho_i$ evolves into $\rho_i'=(1-2d)\rho_i + d I$:
\begin{eqnarray}
&&\Gamma_{(\sigma_3, m=+1, \rho_{0})} \textrm{ occurs with probability $\frac{1-d}{2}$},\nonumber \\
&&\Gamma_{(\sigma_3, m=+1, \rho_{1})}   \textrm{ occurs with probability $\frac{d}{2}$},\nonumber \\
&&\Gamma_{(\sigma_3, m=+1, \rho_{+})}   \textrm{ occurs with probability $\frac{1}{4}$},\nonumber \\
&&\Gamma_{(\sigma_3, m=+1, \rho_{-})}   \textrm{ occurs with probability $\frac{1}{4}$} . \label{probabilitiesofGammas}
\end{eqnarray}
We can also write down the final probe states given that Alice's measurement corresponded to $\sigma_3$ and found the result $m=-1$,
\begin{eqnarray*}
\Gamma_{(\sigma_3, m=-1, \rho_{0})} &=& \ket{\chi_{01}}\bra{\chi_{01}}\\
\Gamma_{(\sigma_3, m=-1, \rho_{1})} &=& \ket{\chi_{11}}\bra{\chi_{11}}\\
\Gamma_{(\sigma_3, m=-1, \rho_{+})} &=& d \ket{\chi_{01}}\bra{\chi_{01}} + (1-d) \ket{\chi_{11}}\bra{\chi_{11}} \\
 && \quad + \sqrt{d(1-d)}\bigl[ \ket{\chi_{01}}\bra{\chi_{11}} + \ket{\chi_{11}}\bra{\chi_{01}} \bigr] \\
\Gamma_{(\sigma_3, m=-1, \rho_{-})} &=& d \ket{\chi_{01}}\bra{\chi_{01}} + (1-d) \ket{\chi_{11}}\bra{\chi_{11}} \\
 && \quad - \sqrt{d(1-d)}\bigl[ \ket{\chi_{01}}\bra{\chi_{11}} + \ket{\chi_{11}}\bra{\chi_{01}} \bigr]  \ ;
\end{eqnarray*}
along with the probabilities for the events to occur which make these the final probe states:
\begin{eqnarray*}
&&\Gamma_{(\sigma_3, m=-1, \rho_{0})} \textrm{ occurs with probability $\frac{d}{2}$},\\
&&\Gamma_{(\sigma_3, m=-1, \rho_{1})}   \textrm{ occurs with probability $\frac{1-d}{2}$},\\
&&\Gamma_{(\sigma_3, m=-1, \rho_{+})}   \textrm{ occurs with probability $\frac{1}{4}$},\\
&&\Gamma_{(\sigma_3, m=-1, \rho_{-})}   \textrm{ occurs with probability $\frac{1}{4}$} .
\end{eqnarray*}

If Alice makes a measurement of $\sigma_3$ and finds the result $m=+1$, Eve's best description of the state of the probe $\Gamma_{(\sigma_3, m=+1)}$, since she is ignorant of the state in which Bob prepared the particle, is described by 
\begin{align*}
&\Gamma_{(\sigma_3, m=+1)}  \\
& = \frac{1-d}{2} \Gamma_{(\sigma_3, +1, \rho_{0})} + \frac{d}{2} \Gamma_{(\sigma_3, m=-1, \rho_{1})} + \frac{1}{4} \Gamma_{(\sigma_3, m=-1, \rho_{+})} + \frac{1}{4} \Gamma_{(\sigma_3, m=-1, \rho_{-})} \\
& = (1-d) \ket{\chi_{00}}\bra{\chi_{00}} + d \ket{\chi_{10}}\bra{\chi_{10}} \ . 
\end{align*}
Similarly, if Alice makes a measurement of $\sigma_3$ and finds the result $m=-1$ then Eve's best description of the state of the probe system is
\[
\Gamma_{(\sigma_3, m=-1)} = d \ket{\chi_{01}}\bra{\chi_{01}} + (1-d) \ket{\chi_{11}}\bra{\chi_{11}} \ . 
\]

So the task that Eve is left with, once she learns that Alice made the measurement corresponding to $\sigma_3$, is to distinguish between these two equiprobable probe output states, $\Gamma_{(\sigma_3, +1)}$ and $\Gamma_{(\sigma_3, -1)}$.   But these two final probe states are related to the states $\Gamma_0$ and $\Gamma_1$ by a unitary transformation.  For the unitary operator 
\[
V  = (1-2d)\bigl( \ket{\eta_{1}}\bra{\eta_{1}} + \ket{\eta_{3}}\bra{\eta_{3}} \bigr)  +  2\sqrt{d(1-d)}\bigl( \ket{\eta_{1}}\bra{\eta_{3}} + \ket{\eta_{3}}\bra{\eta_{1}} \bigr) \ ,
\] 
\begin{align*}
\Gamma_{(\sigma_3, +1)} &= V \Gamma_0 V^\dagger \\
\Gamma_{(\sigma_3, +1)} &= V \Gamma_1 V^\dagger \ .
\end{align*}
The measurement that optimally distinguishes between these two equiprobable mixed states is $\{ VE_1 V^\dagger, VE_2V^\dagger, VE_3V^\dagger, VE_4V^\dagger \}$, where the $E_i$'s are the operators that are used by Fuchs \textit{et al} to optimally distinguish between the states $\Gamma_0$ and $\Gamma_1$.  
The probability for the four different possible measurement outcomes that Eve may get, given that the probe is in the state $\Gamma_{(\sigma_3, +1)}= V \Gamma_0 V^\dagger$ or $\Gamma_{(\sigma_3, -1)} = V\Gamma_1 V^\dagger$, are the same as those shown in Table \ref{distinguishing}.  We can again apply Bayes' rule $\Pr(\Gamma_{(\sigma_3, \pm1)}|E_i) = \half \Pr(E_i | \Gamma_{(\sigma_3, \pm1)})/ \Pr(E_i)$ to determine these probabilities that arise in the calculation of the mutual information.  

The same thing occurs when Alice's measurement corresponds to $\sigma_1$, due to the symmetry of the probe and the final state of the particle as it reaches Alice.  That is, the evolution of the probe system requires Eve to distinguish between two different mixed states of the probe system,
\[
\Gamma_{(\sigma_1, +1)} = (1-d) \ket{\chi_{++}}\bra{\chi_{++}} + d \ket{\chi_{-+}}\bra{\chi_{-+}}\]
and
\[
\Gamma_{(\sigma_1, -1)} = d \ket{\chi_{+-}}\bra{\chi_{+-}} + (1-d) \ket{\chi_{--}}\bra{\chi_{--}} \ . 
\]
Furthermore, there exists a unitary operator $W$ so that the measurement described by the operators $\{WF_1W^\dagger, WF_2W^\dagger, WF_3W^\dagger, WF_4W^\dagger \}$  is the best for  distinguishing between these two mixed states.  Furthermore, the probabilities of the various outcomes match those that occurred when Alice made the measurement corresponding to $\sigma_3$.

This proof is dependent upon two points:  First, that the requirements imposed by our implementation of the symmetric probe do not hinder the generality of this analysis.  Second, that the measurement found by Fuchs \textit{et al} is the best way to distinguish between two equiprobable mixed states $\Gamma_0$ and $\Gamma_1$, and therefore the related measurement can best distinguish between the equiprobable states $\Gamma_{(\sigma_3, +1)}= V \Gamma_0 V^\dagger$ and $\Gamma_{(\sigma_3, -1)} = V\Gamma_1 V^\dagger$.  This concludes the proof on the optimal eavesdropping activity.

\end{document}